\documentclass[11pt]{article}      
\pdfoutput=1
\usepackage{amssymb}
\usepackage{amsmath}
\usepackage{amsthm}
\usepackage[]{graphicx}
\usepackage[]{subfigure}
\usepackage{tensor}
\usepackage{float}
\usepackage{color}
\usepackage{cancel}
\usepackage{setspace}
\usepackage{fancyhdr}
\usepackage{framed}
\usepackage{braket}
\usepackage{tikz}
\usepackage{comment}
\usepackage{bbold}
\usepackage{bbm}

\usepackage[noadjust]{cite}
\let\OLDthebibliography\thebibliography
\renewcommand\thebibliography[1]{
  \OLDthebibliography{#1}
  \setlength{\parskip}{0pt}
  \setlength{\itemsep}{0pt plus 0.3ex}
}

\usepackage[textwidth=6.5in,top=1.2in,bottom=1.2in]{geometry}

\usepackage{setspace}
\usepackage{simplewick}
\usepackage{float}
\usepackage{mathtools}
\usepackage{blkarray, bigstrut} %
\usepackage[bookmarks,linktocpage, colorlinks=true, plainpages = false, citecolor = red,  linkcolor=blue, urlcolor = magenta, filecolor = blue]{hyperref} 
\usepackage{braket}
\usepackage{mathrsfs}
\usepackage{slashed}
\usepackage[title]{appendix}
\numberwithin{equation}{section}

\begin{document}

\thispagestyle{empty}
\nopagebreak

\title{\begin{center}\bf Revisiting the Schr\"odinger-Dirac equation\end{center}}

\vfill
\author{Nicolas\,Fleury$^1$\footnote{\href{mailto:psadeghi20@ubishops.ca}{nfleury22@ubishops.ca}},\, Fay\c{c}al\,Hammad$^{1,2}$\footnote{\href{mailto:fhammad@ubishops.ca}{fhammad@ubishops.ca}},\;\,Parvaneh\,Sadeghi$^1$\footnote{\href{mailto:psadeghi20@ubishops.ca}{psadeghi20@ubishops.ca}}}
\date{ }
\maketitle

\begin{center}
	\vspace{-0.7cm}
{\it  $\,^1$Department of Physics and Astronomy, Bishop's University, 2600 College Street, Sherbrooke, QC, J1M~1Z7 Canada}\\
	{\it  $\,^2$Physics Department, Champlain 
College-Lennoxville, 2580 College Street, Sherbrooke,  
QC, J1M~2K3 Canada}
\end{center}
\bigskip
\abstract{In flat spacetime, the Dirac equation is the ``square root'' of the Klein-Gordon equation in the sense that by applying the square of the Dirac operator to the Dirac spinor, one recovers the Klein-Gordon equation duplicated for each component of the spinor. In the presence of gravity, applying the square of the curved-spacetime Dirac operator to the Dirac spinor does not yield the curved-spacetime Klein-Gordon equation, but yields, instead, the Schr\"odinger-Dirac covariant equation. First, we show that the latter equation gives rise to a generalization to spinors of the covariant Gross-Pitaevskii equation. Next, we show that while the Schr\"odinger-Dirac equation is not conformally invariant, there exists a generalization of the equation that is conformally invariant but which requires a different conformal transformation of the spinor than the one required by the Dirac equation. The new conformal factor acquired by the spinor is found to be a matrix-valued factor obeying a differential equation that involves the Fock-Ivanenko line element. The Schr\"odinger-Dirac equation coupled to the Maxwell field is then revisited and generalized to particles with higher electric and magnetic moments while respecting gauge symmetry. Finally, Lichnerowicz's vanishing theorem in the conformal frame is also discussed.}

\newpage
\section{Introduction}\label{sec:Intro}
It is well known that, historically \cite{Dirac}, Dirac derived his equation by seeking a first-order relativistic covariant equation of the Schr\"odinger form for the wavefunction $\psi$. In such a process, one demands that the wave equation describe a relativistic free particle of four-momentum $p^a$ and of mass $m$, such that the usual relativistic energy-momentum relation $\eta_{ab}p^a p^b-m^2=0$ hold\footnote{We set throughout the paper $G=\hbar=c=1$, and we work with the metric signature $(+,-,-,-)$. We shall also use throughout the paper Latin letters to denote tangent-space and flat-spacetime indices. We reserve Greek letters for curved-spacetime indices.}. Here, $\eta_{ab}$ is the Minkowski metric of flat spacetime and $\eta^{ab}$ is its inverse. To fulfill such a requirement, every single component of the 4-component wavefunction $\psi$, now called a spinor field, should obey the Klein-Gordon equation that describes spin-$0$ fields. Indeed, it is well known (see e.g., Ref.\,\cite{Greiner}) that applying the operator $i\gamma^a\partial_a+m$ from the left to the first-order Dirac equation $\left(i\gamma^a\partial_a-m\right)\psi=0$, where $\gamma^a$ are the Dirac gamma matrices, yields the flat-spacetime second-order Klein-Gordon equation $(\eta^{ab}\partial_a\partial_a+m^2)\,\psi=0$. The reason why the Klein-Gordon operator $\eta^{ab}\partial_a\partial_b+m^2$ acts on every single component of the four-spinor $\psi$ is due to the fact that such an operator is a diagonal operator. 

Moving on to curved spacetimes, \textit{i.e.}, in the presence of gravity, this simple link between the Klein-Gordon equation and the Dirac equation breaks down. While both equations keep their respective forms thanks to the principle of covariance, one does not recover the curved-spacetime Klein-Gordon equation by starting from the curved-spacetime Dirac equation. What one obtains, instead, is a covariant equation, first derived by Schr\"odinger \cite{Schrodinger,SchrodingerNote}, that we shall call here the ``Schr\"odinger-Dirac'' equation for reasons to be explained in Sec.\,\ref{sec:SchroDirac}. The latter equation contains several extra terms besides those contained in the curved-spacetime Klein-Gordon equation, and it consists of four coupled differential equations. This is in contrast to what one might expect since one would naturally demand again that the covariant energy-momentum relation $g^{\mu\nu}p_\mu p_\nu-m^2=0$ for the particle be satisfied in a spacetime of metric $g_{\mu\nu}$. The latter relation, when written in a covariant operator form thanks to the replacement $p_\mu\rightarrow i\nabla_\mu$, does indeed give rise to the curved-spacetime Klein-Gordon equation for a scalar field $\phi$ without any extra term, \textit{i.e.}, $(g^{\mu\nu}\nabla_\mu\nabla_\nu+m^2)\,\phi=0$.

The mathematical reason why such a simple link between the Dirac equation and the Klein-Gordon equation does not hold in curved spacetimes is that the non-commuting gamma matrices become spacetime-dependent. As a result, the second derivative operator also acts on those matrices from the left, giving rise to extra non-diagonal terms in the final equation. The physical reason why such a simple link does not hold in the presence of gravity is because the latter couples the different components of the four-spinor. This prevents those components from obeying independent second-order differential equations as they do in flat spacetime. It turns out, as we shall see in this paper, that the non-minimal coupling with gravity makes the equation take the same nonlinear form as the more familiar Gross-Pitaevskii equation in curved spacetime \cite{Gross,Pitaevskii,Anandan,Suarez}. To the best of our knowledge, this has not been pointed out in the literature before.

On the other hand, it is also a known fact that when conformally deforming spacetime the Dirac equation remains conformally invariant whereas one easily shows that the Schr\"odinger-Dirac equation does not. When recalling that the Klein-Gordon equation in curved spacetime is not conformally invariant but its non-minimally coupled version is \cite{What?}, it becomes of great interest to seek a generalization of the already non-minimally coupled Schr\"odinger-Dirac equation that would also be conformally invariant. We show in this paper that such a generalization does indeed exist and that it requires the spinor field to conformally transform by bringing in a conformal factor that is different from the one required by the Dirac equation. Indeed, we show that the conformal factor the spinor comes with is a matrix-valued function that obeys in the conformal frame a simple differential equation involving the Fock-Ivanenko line element in a fundamental way.

Besides deriving his equation for neutral spinors, Schr\"odinger also showed in the same paper that in the presence of the Maxwell field, one also extracts a second-order differential equation from the curved-spacetime Dirac equation minimally coupled to the Maxwell field \cite{Schrodinger}. The resulting equation displays the correct gyromagnetic ratio of the particle \cite{Pagels} just as the more familiar non-relativistic flat-spacetime Pauli-Schr\"odinger equation does. We show in this paper that when starting from the curved-spacetime Dirac equation that contains, besides the usual minimal-coupling term, an anomalous magnetic moment term the resulting second-order differential equation generalizes the Schr\"odinger-Dirac equation to particles with an anomalous magnetic moment. The physical meaning of the resulting extra terms in the equation is discussed. Afterwards, we generalize further the Schr\"odinger-Dirac equation to describe particles with higher electric and magnetic moments. We show that the resulting equation displays correction terms for the magnetic moment of the particle that are proportional to spacetime curvature terms and their derivatives. Those terms are very similar to the more familiar terms that arise in quantum field theories from the calculation of the expectation value of the stress-energy momentum tensor in curved spacetime. We shall discuss the origin of this coincidence in great detail. 

Finally, since the well-known Lichnerowicz vanishing theorem \cite{Lichnerowicz} is  based on the Lichnerowicz-Schr\"odinger operator, which is the operator acting on the spinor $\psi$ in the massless Schr\"odinger-Dirac equation, we examine the fate of the theorem in the conformal frame. We show that one reaches different conclusions about the link between harmonic spinors and the curvature of the spin manifold depending on whether one relies on Lichnerowicz's identity in the conformal frame or one relies on the conformally transformed identity.

The reminder of this paper is structured as follows. In Sec.\,\ref{sec:SchroDirac}, we briefly review the way the Schr\"odinger-Dirac equation is obtained, then we extract from it a nonlinear equation for spinors which is the analog of the nonlinear covariant Gross-Pitaevskii equation for scalars in curved spacetimes. In Sec.\,\ref{sec:ConfSchroDirac}, we first derive in detail the conformal noninvariance of the Schr\"odinger-Dirac equation and point out some subtleties behind the derivation. We then build a conformally invariant version of the equation and examine the nature and behavior of the new conformal factor. In Sec.\,\ref{sec:SchroDiMaxwell}, we revisit the Schr\"odinger-Dirac equation coupled to the Maxwell field and derive a generalized equation for particles with an anomalous magnetic moment as well as for particles with higher electric and magnetic moments. In Sec.\,\ref{sec:Lichnerowicz}, we revisit Lichnerowicz's vanishing theorem for conformally transformed spin manifolds. We conclude this paper with a brief summary and conclusion section.
\section{\!\!\!The Schr\"odinger-Dirac equation and the curved-spacetime Gross-Pitaevskii equation}\label{sec:SchroDirac}
We first briefly recall in this section the derivation of the covariant Schr\"odinger-Dirac equation, and then show the interesting emergence from it of a curved-spacetime Gross-Pitaevskii-like equation.

The generalization of the Dirac equation to a curved spacetime of metric $g_{\mu\nu}$ is accomplished thanks to the use of the spacetime vierbeins $e^a_\mu$ \cite{Weyl1,Weyl2,FockIvanenko1,FockIvanenko2,Fock}, defined by $\eta_{ab}e^a_\mu e^b_\nu=g_{\mu\nu}$ \cite{RicciCivita}. The inverse $e^\mu_a$ of the vierbeins are analogously defined by $g_{\mu\nu}e^\mu_ae^\nu_b=\eta_{ab}$. The vierbeins allow one to build the curved-spacetime gamma matrices $\gamma^\mu$ via the contraction $\gamma^\mu=e^\mu_a\gamma^a$, as a consequence of which the usual flat-spacetime anti-commutation relations of the gamma matrices, $\{\gamma_a,\gamma_b\}=2\eta_{ab}$, become $\{\gamma_\mu,\gamma_\nu\}=2g_{\mu\nu}$ in curved spacetimes \cite{Tetrode}. Furthermore, the partial derivatives $\partial_a$ should be replaced by the spin-covariant derivatives $D_\mu$ defined by $D_\mu=\partial_\mu+\Lambda_\mu$, where the Fock-Ivanenko coefficients $\Lambda_\mu$ involve both the spin connection $\omega_\mu^{\,ab}$ and the gamma matrices according to $\Lambda_\mu=\frac{1}{8}\,\omega_\mu^{\,ab}[\gamma_a,\gamma_b]\equiv\omega_\mu^{\,ab}\Sigma_{ab}$. We have introduced here, for convenience, the useful symbol $\Sigma_{ab}$, also called spin tensor, to stand for the often-reoccurring commutator $\frac{1}{8}[\gamma_a,\gamma_b]$\footnote{In the literature, the spin tensor is sometimes defined slightly differently as, $\sigma_{ab}=\tfrac{i}{2}[\gamma_a,\gamma_b]$.} \cite{QFT1,QFT2}.

The spin connection $\omega_\mu^{\,ab}$ is related to the vierbeins and the Christoffel symbols $\Gamma_{\mu\nu}^\lambda$ by the usual defining relation $\omega_\mu^{\,\,ab}=e^a_\nu\partial_\mu e^{\nu b}+\Gamma_{\mu\nu}^\lambda e^{a}_\lambda e^{\nu b}$. To distinguish the covariant derivative operator $D_\mu$ acting on spinors from the usual covariant derivative operator acting on tensors, we denote the latter operator by the symbol $\nabla_\mu$. We reserve the symbol $\mathcal{D}_\mu$ for the total covariant derivative operator acting on objects that carry both curved-space and tangent-space/spinor indices. With these ingredients, the curved-spacetime Dirac equation takes the form $(i\gamma^\mu D_\mu-m)\,\psi=0$. Applying the operator $i\gamma^\mu D_\mu+m$ from the left to the latter equation, yields the covariant Schr\"odinger-Dirac equation \cite{Schrodinger,SchrodingerNote}:
\begin{equation}\label{SchroDirac}
    \left(g^{\mu\nu}\mathcal{D}_\mu \mathcal{D}_\nu+m^2+\tfrac{1}{4}R\right)\psi=0.
\end{equation}
The Ricci scalar $R$ in this equation arises thanks to the emergence of the square of the covariant derivative operator $\slashed{D}\equiv\gamma^\mu D_\mu$. This equation was first derived by Schr\"odinger, whence the name we chose for this equation. The detailed derivation of this equation, as well as subtleties concerning such a derivation are given in Appendix \ref{sec:AppDerivation}. To the best of our knowledge, such subtleties have not been pointed out before. By comparing Eq.\,(\ref{SchroDirac}) to the curved-spacetime Klein-Gordon equation, $(g^{\mu\nu}\nabla_\mu\nabla_\nu+m^2)\,\phi=0$, it is clear that not a single component of the four-spinor $\psi$ obeys the latter equation as they all do in flat spacetime. The reason is that the operator $g^{\mu\nu}\mathcal{D}_\mu \mathcal{D}_\nu$ in Eq.\,(\ref{SchroDirac}) is a non-diagonal matrix operator.

In Ref.\,\cite{SchrodingerNote}, it was suggested that one ``ought'' to call Eq.\,(\ref{SchroDirac}) the Schr\"odinger-Lichnerowicz formula since Lichnerowicz, independently in Ref.\,\cite{Lichnerowicz}, re-derived the formula $\slashed{D}^2=g^{\mu\nu}\mathcal{D}_\mu \mathcal{D}_\nu+\frac{1}{4}R$ that leads to the equation. We shall indeed call here the latter formula the Lichnerowicz-Schr\"odinger identity even though it is sometimes also called the Bochner-Weitzenb\"ok identity in reference to other related versions of the identity in the mathematics literature \cite{SpinGeometry,Li}. However, it is clear that Eq.\,(\ref{SchroDirac}) is neither merely a formula, nor does it reduce to an identity between operators. It is a dynamical equation governing the spacetime evolution of a spin-$\frac{1}{2}$ field. We actually chose for it such a name in analogy to the name chosen for the non-relativistic Pauli-Schr\"odinger equation\footnote{Note that in Ref.\,\cite{ChapmanCerceau} (and in Ref.\,\cite{QuantumPhase}), Eq.\,(\ref{SchroDirac}) was called the ``generalized covariant Pauli-Schr\"odinger equation''. We do not find such a name adequate, for Eq.\,(\ref{SchroDirac}) is not merely a covariant generalization of the Pauli-Schr\"odinger equation as the latter is neither a special-relativistic equation nor an equation describing four-spinors. Another possible name for the equation would be ``modified Klein-Gordon equation'' given in Ref.\,\cite{Pollock}. We do not find such a name adequate either, for Eq.\,(\ref{SchroDirac}) is not merely a modification of an equation that describes a scalar field. Eq.\,(\ref{SchroDirac}) describes a completely different physical entity and requires different mathematical objects and tools.}. The latter equation was postulated by Pauli \cite{Pauli1} who extended the non-relativistic Schr\"odinger equation to non-relativistic spin-$\frac{1}{2}$ particles. Similarly, Eq.\,(\ref{SchroDirac}) was {\it derived} by Schr\"odinger who extracted it from the curved-spacetime Dirac equation as a second-order differential equation that turned out to involve the Ricci scalar and, yet, maintain a matrix form as it is the case with the original Dirac equation.

It is worth noting here that, unlike the Dirac equation in curved spacetime, the Schr\"odinger-Dirac equation (\ref{SchroDirac}) does not mix the chiral components of the spinor field $\psi$. Indeed, by applying the chiral projectors $P_{L,R}=\frac{1}{2}\left(1\mp\gamma^5\right)$ to Eq.\,(\ref{SchroDirac}) from the left, the chiral components $\psi_{L,R}$ get decoupled and obey separately the same equation:
\begin{equation}\label{SchroDiracLeftRight}
    \left(g^{\mu\nu}\mathcal{D}_\mu \mathcal{D}_\nu+m^2+\tfrac{1}{4}R\right)\psi_{L,R}=0.
\end{equation}
Use has been made here of the fact that the matrix $\gamma^5$ commutes with the spinor tensor $\Sigma_{ab}$ hiding inside the derivative operators $\mathcal{D}_{\mu}$ and $\mathcal{D}_{\nu}$.

After this brief introduction to the equation and the tools it requires, we shall discuss now some interesting physics that emerges from the equation. In fact, although Eq.\,(\ref{SchroDirac}) seems to be only of academic interest since the $R$ term is orders of magnitude smaller than the mass term in that equation (as Schr\"odinger himself pointed it out \cite{Schrodinger}), the equation is actually very rich in physical content. It turns out, indeed, that the coupling of the spinor field $\psi$ to the Ricci scalar in Eq.\,(\ref{SchroDirac}) offers a novel possibility that is not found even in the curved spacetime Klein-Gordon equation. 

Physically, Eq.\,(\ref{SchroDirac}) means that the Dirac spinor is actually indirectly coupled to other forms of matter if there are any in that region of spacetime, for the Ricci scalar is then determined by matter distribution via Einstein's field equations. However, the presence of the spin-$\frac{1}{2}$ field itself is already a source for gravitation even in the absence of any other forms of matter. Indeed, contracting both sides of the Einstein field equations $R_{\mu\nu}-\frac{1}{2}g_{\mu\nu}=-8\pi T_{\mu\nu}$ with the inverse metric tensor $g^{\mu\nu}$, leads to $R=8\pi T$, where $T_{\mu\nu}$ is the energy-momentum tensor of matter and $T=g^{\mu\nu}T_{\mu\nu}$ is its trace. On the other hand, the energy-momentum tensor of the Dirac field in curved spacetime is $\frac{i}{2}[\bar{\psi}\gamma_{(\mu}D_{\nu)}\psi-(D_{(\mu}\bar{\psi})\gamma_{\nu)}\psi]$ \cite{QFT1}, where parentheses around two indices stand for symmetrization in those indices and $\bar{\psi}=\psi^\dagger\gamma^0\psi$. The trace $T$ of this energy-momentum tensor is then easily evaluated to be $T=m\,\bar{\psi}\psi$, where we have used both the Dirac equation and its Dirac adjoint. This implies then that the Ricci scalar $R$ is given by $R=8\pi m\,\bar{\psi}\psi$. Substituting this into Eq.\,(\ref{SchroDirac}), the latter takes the following nonlinear form:
\begin{equation}\label{SchroDiracNonLinear}
\left(g^{\mu\nu}\mathcal{D}_\mu\mathcal{D}_\nu+m^2+2\pi m\,\bar{\psi}\psi\right)\psi=0.
\end{equation}
Remarkably, this equation is very reminiscent of the nonlinear Gross-Pitaevskii equation used to describe superfluids \cite{Gross,Pitaevskii}. Equation (\ref{SchroDiracNonLinear}) can be seen as a generalization to spinor fields of the covariant Gross-Pitaevskii equation that describes scalar fields in curved spacetime \cite{Anandan,Suarez}. In other words, classical gravity is able to induce a superfluid-like dynamics in spin-$\frac{1}{2}$ quantum fields thanks to the back-reaction of the latter on the background spacetime. 

It is interesting to note here that in Ref.\,\cite{Heisenberg1} Heisenberg introduced the nonlinear term $(\bar{\psi}\psi)\psi$ as a \textit{postulated} new term on the right-hand side of the \textit{Dirac equation}, and he has even quantized the resulting nonlinear wave equation \cite{Heisenberg2,Heisenberg3}. It should be emphasized, though, that whereas Heisenberg postulated such a nonlinear term for the Dirac equation, that term arose here on the left-hand side of the Schr\"odinger-Dirac equation from \textit{first principles}. Furthermore, whereas Heisenberg postulated the term in the hope of unifying the wave equations of matter and explaining the origin of the electron charge \cite{Heisenberg1}, the term emerges here simply as a consequence of the universal coupling of matter to gravity.

Restoring the physical constants to Eq.\,(\ref{SchroDiracNonLinear}) and comparing its nonlinear term with the term $\frac{2m}{\hbar^2} g\,|\varphi|^2\varphi$ of the covariant Gross-Pitaevskii equation \cite{Anandan}, where $g$ is related to the scattering length $a_s$ of the bosons by $g=4\pi\hbar^2 a_s/m$, we conclude that the analog of the scattering length $a_s$ in the fermion case is $mG/4c^2$; which is an extremely small quantity, of course. It is important to emphasize here that this observation has never been pointed out in the literature before, for what Schr\"odinger found attractive about the $R$ term in his equation was rather the possibility of interpreting it as a kind of a mass-generating term \,\cite{Schrodinger}. A similar argument to Schr\"odinger's is also found in Ref.\,\cite{Pagels}. 

It is worth noting here also that, unlike the Schr\"odinger-Dirac equation (\ref{SchroDirac}), the nonlinear equation (\ref{SchroDiracNonLinear}) does mix the chiral components of the spinor, for we have $\bar{\psi}\psi=\bar{\psi}_R\psi_L+\bar{\psi}_L\psi_R$.
\section{A modified Schr\"odinger-Dirac equation}\label{sec:ConfSchroDirac}
Looking at the form of Eq.\,(\ref{SchroDirac}), one cannot help but think of the conformally invariant version of the non-minimally coupled Klein-Gordon equation in curved spacetime: $(g^{\mu\nu}\nabla_\mu\nabla_\nu+m^2+\frac{1}{6}R)\,\phi=0$. This equation is conformally invariant only because of the presence of the specific factor $\frac{1}{6}$ in front of the Ricci scalar $R$. Any other factor in front of $R$ in the latter equation would not render it conformally invariant \cite{What?}. One might then naturally expect that while the Schr\"odinger-Dirac equation (\ref{SchroDirac}) is not conformally invariant, a simply different numerical factor in front of $R$ in that equation could render the latter conformally invariant. However, it turns out that things are more subtle and interesting as we shall see. Before working out the conformal transformation of Eq.\,(\ref{SchroDirac}), let us first recall what we mean by a conformal transformation, a conformal frame and conformal invariance. 

What we mean here by a conformal transformation ---\,also known as a Weyl conformal transformation in order to distinguish it from the conformal coordinate transformations\,--- is the spacetime-dependent rescaling of the metric $g_{\mu\nu}$. In other words, one builds a conformal spacetime (also called a conformal frame) of metric $\tilde{g}_{\mu\nu}$ by simply multiplying the original spacetime metric $g_{\mu\nu}$ pointwise by a positive and everywhere regular spacetime-dependent factor. Such a factor is usually denoted by $\Omega^2(x)$, such that $\tilde{g}_{\mu\nu}=\Omega^2g_{\mu\nu}$ (see, e.g., Ref.\,\cite{Wald,Conformal}). 

Such a transformation of the metric also transforms the mass $m$ into $\tilde{m}=\Omega^{-1}m$ and the spinor $\psi$ into $\tilde{\psi}=\Omega^{-\frac{3}{2}}\psi$. Furthermore, by using the link between the vierbeins and the metric, we also learn that the old vierbeins $e^a_\mu$ are transformed into $\tilde{e}^a_\mu=\Omega\,e^a_\mu$. In addition, since the Christoffel symbols are then transformed into $\tilde{\Gamma}_{\mu\nu}^\lambda$, the spin connection is also transformed into $\tilde{\omega}_\mu^{\,ab}$ and the Fock-Ivanenko coefficient is transformed into $\tilde{\Lambda}_\mu$. With these transformed terms, one easily derives the expression of the transformed version of the covariant derivative operator $\tilde{D}_\mu$ as well as the transformed Ricci scalar $\tilde{R}$, the explicit expressions of which are all given in Eq.\,(\ref{ConformalConnections}) of Appendix \ref{sec:AppConfDirac}. With all these transformed mathematical objects, the Dirac equation in the conformal spacetime turns out to have exactly the same form it has in the original spacetime (see Eq.\,(\ref{ProofConformalDirac}) for the detailed steps of the derivation): \begin{equation}\label{ConformalDirac}
(i\tilde{\gamma}^\mu\tilde{D}_\mu-\tilde{m})\,\tilde{\psi}=0.
\end{equation}
Equations that preserve their form under a conformal transformation are called conformally invariant equations\footnote{It is worth noting that investigating the effect of a Weyl transformation on the various equations of physics can be more than just a formal check of conformal (non)-invariance of equations. When properly interpreted, the results help one gain novel insights about the {\it nature} of the link between physical concepts and entities. Implementing such a philosophy with such a specific goal in mind has indeed been fruitful in multiple recent works, ranging from the physics of quasi-local masses \cite{MS,HH} to the physics of wormholes and black holes \cite{BHWormhole,BH,ST,ParallelBH}. In the domain of quantum physics, which is the case here, and contrary to one's expectation, a mundane noninvariance of an equation could even shed new light on the interpretation issue emerging in the foundations of quantum mechanics \cite{What?}.}.

Now, thanks to the conformal invariance of the Dirac equation, it is straightforward to see what form the Schr\"odinger-Dirac equation will have in the conformal frame if ever one starts from Eq.\,(\ref{ConformalDirac}). One simply needs, indeed, to apply the operator $\tilde{\gamma}^\mu\tilde{D}_\mu$ from the left to Eq.\,(\ref{ConformalDirac}). The terms such a procedure has led to based on the metric $g_{\mu\nu}$ do in fact necessarily emerge here unaltered when based on the metric $\tilde{g}_{\mu\nu}$, only to be decorated everywhere by tildes. In addition, however, one extra term arises due to the position-dependent mass $\tilde{m}$. In fact, applying the operator $\tilde{\gamma}^\mu\tilde{D}_\mu$ to the left-hand side of Eq.\,(\ref{ConformalDirac}) the conformally transformed Schr\"odinger-Dirac equation takes the following form:
\begin{equation}\label{ConformalSchroDiracI}
\left(\tilde{g}^{\mu\nu}\tilde{\mathcal{D}}_\mu\tilde{\mathcal{D}}_\nu+\tilde{m}^2+\tfrac{1}{4}\tilde{R}\right)\,\tilde{\psi}=i \tilde{m}\frac{\Omega_{,\mu}}{\Omega}\tilde{\gamma}^\mu\tilde{\psi}.
\end{equation}
It is thus clear that the equation is conformally noninvariant and that the transformed spinor field $\tilde{\psi}$ would obey in the conformal spacetime a slightly different dynamics from the one it obeyed in the original spacetime due to the single extra term on the right-hand side of Eq.\,(\ref{ConformalSchroDiracI}). Only massless spinors would lead to the same equation in the conformal frame.

The other way of searching for the conformal version of the Schr\"odinger-Dirac equation is to start from the latter as given by Eq.\,(\ref{SchroDirac}), which we know holds in the original spacetime, and then conformally transform all the terms of that equation. The detailed calculations are given in Appendix \ref{sec:AppConfSchrodirac}, and the final result is the following new equation:
\begin{equation}
\label{ConformalSchroDiracII}
\left(\tilde{g}^{\mu\nu}\tilde{\mathcal{D}}_\mu\tilde{\mathcal{D}}_\nu+\tilde{m}^2+\tfrac{1}{4}\tilde{R}\right)\,\tilde{\psi}=-\frac{\Omega_{,\mu}}{\Omega}\tilde{\gamma}^\mu\tilde{\gamma}^\nu\tilde{D}_\nu\tilde{\psi}.
\end{equation}
We clearly see that only when one assumes the Dirac equation holds in the conformal spacetime, \textit{i.e.}, only when $i\tilde{\gamma}^\mu\tilde{D}_\mu\tilde{\psi}=\tilde{m}\tilde{\psi}$, does the right-hand side of Eq.\,(\ref{ConformalSchroDiracII}) coincide with the right-hand side of  Eq.\,(\ref{ConformalSchroDiracI}). The Schr\"odinger-Dirac equation is thus conformally invariant for massless spinors, but only if the latter are \textit{also assumed} to obey the Dirac equation in the conformal frame. It is worth noting here that the last term on the right-hand side of Eq.\,(\ref{ConformalSchroDiracII}) is a generalization to spinors of a similar extra term arising on the right-hand side of the conformally transformed curved-spacetime Klein-Gordon equation. The corresponding term for the latter equation in the conformal frame is $\frac{2\Omega_{,\mu}}{\Omega}\tilde{\nabla}^\mu\tilde{\phi}$ \cite{What?}.

However, the non-minimally coupled Klein-Gordon equation does not contain any extra term when moving to the conformal frame. We are therefore naturally led to look for a modified Schr\"odinger-Dirac equation that would be conformally invariant in analogy with the non-minimally coupled Klein-Gordon equation. As the conformal invariance of the latter is achieved by the mere presence of the term $\frac{1}{6}R\phi$ on the left-hand side of the equation, a natural guess for a modified Schr\"odinger-Dirac equation is $\left(g^{\mu\nu}\mathcal{D}_\mu\mathcal{D}_\nu+m^2+\xi R\right)\,\psi=0$ for some numerical factor $\xi$. 

Starting from such a guess for the modified equation, we assume the following more general conformal transformation of the spinor field,
\begin{equation}\label{GenConfPsi}
\psi(x)=\Omega^{\frac{3}{2}}\mathbb{S}(\Omega,x)\tilde{\psi}(x),
\end{equation}
where $\mathbb{S}(\Omega,x)$ is an unknown (possibly an invertible $4\times4$ matrix-valued) functional of the conformal factor $\Omega(x)$. Inserting this expression into our guess for the modified Schr\"odinger-Dirac equation, the same steps followed in Eq.\,(\ref{ConfSchroDiracTransformation}) for the usual Schr\"odinger-Dirac equation yield now,
\begin{multline}\label{Modified1}
   \mathbb{S}\left(\tilde{g}^{\mu\nu}\tilde{\mathcal{D}}_\mu\tilde{\mathcal{D}}_\nu+\tilde{m}^2+\xi \tilde{R}\right)\tilde{\psi}+2\left[\tilde{\mathcal{D}}^\mu\mathbb{S}+\frac{\Omega_{,\nu}}{2\Omega}\tilde{\gamma}^\nu\tilde{\gamma}^\mu\mathbb{S}\right]\tilde{D}_\mu\tilde{\psi}\\
+\left[\tilde{\mathcal{D}}^\mu\tilde{\mathcal{D}}_\mu\mathbb{S}+\left(\tfrac{3}{2}-6\xi\right)\frac{\tilde{\Box}\Omega}{\Omega}\mathbb{S}-\left(3-12\xi\right)\frac{\tilde{g}^{\mu\nu}\Omega_{,\mu}\Omega_{,\nu}}{\Omega^2}\mathbb{S}+\frac{\Omega_{,\nu}}{\Omega}\tilde{\gamma}^{\nu}\tilde{\gamma}^\mu\tilde{\mathcal{D}}_\mu\mathbb{S}\vphantom{\frac{\tilde{\Box}\Omega}{\Omega}}\right]\tilde{\psi}=0.
\end{multline}
Therefore, for the modified Schr\"odinger-Dirac equation to be conformally invariant the content of each of the pairs of square brackets in Eq.\,(\ref{Modified1}) has to vanish. The vanishing of the content of the first pair of square brackets that multiplies the term $\tilde{D}_\mu\tilde{\psi}$ leads to the following constraint on the functional $\mathbb{S}(\Omega,x)$:
\begin{equation}\label{Constraint}
    \tilde{\mathcal{D}}_\mu\mathbb{S}=-\frac{\Omega_{,\nu}}{2\Omega}\tilde{\gamma}^\nu\tilde{\gamma}_\mu\mathbb{S}.
\end{equation}
This equation shows that the functional $\mathbb{S}(\Omega,x)$ \textit{cannot} be a scalar, but only a matrix-valued functional of the conformal factor $\Omega$. Contracting both sides of Eq.\,(\ref{Constraint}) by $\tilde{\gamma}^\mu$, and applying the derivative operator $\tilde{\mathcal{D}}^\mu$ to both sides of Eq.\,(\ref{Constraint}) lead to the following two equations, respectively: 
\begin{equation}\label{2OtherConstraints}
\tilde{\gamma}^\mu\tilde{\mathcal{D}}_\mu\mathbb{S}=\frac{\Omega_{,\nu}}{\Omega}\tilde{\gamma}^\nu\mathbb{S},\qquad    \tilde{\mathcal{D}}^\mu\tilde{\mathcal{D}}_\mu\mathbb{S}=-\frac{\tilde{\Box}\Omega}{2\Omega}\mathbb{S}.
\end{equation} 
Inserting these two constraints on $\mathbb{S}(\Omega,x)$ into the content of the second pair of square brackets that multiplies $\tilde{\psi}$ in Eq.\,(\ref{Modified1}), and setting the resulting expression equal to zero, straightforwardly yields the numerical value of $\xi$ to be $\frac{1}{6}$. Remarkably, this value coincides exactly with the numerical value of the factor multiplying $R$ in the conformally invariant non-minimally coupled Klein-Gordon equation \cite{What?}. The modified Schr\"odinger-Dirac equation that is conformally invariant then reads
\begin{equation}\label{ModifiedSD}
\left(g^{\mu\nu}\mathcal{D}_\mu\mathcal{D}_\nu+m^2+\tfrac{1}{6} R\right)\,\psi=0,
\end{equation}
in which the spinor $\psi$ transforms as $\tilde{\psi}=\Omega^{-\frac{3}{2}}\mathbb{S}^{-1}(\Omega,x)\psi$, where the matrix-valued functional $\mathbb{S}(\Omega,x)$ obeys all three differential equations as given in Eqs.\,(\ref{Constraint}) and (\ref{2OtherConstraints}). Furthermore, by multiplying both sides of Eq.\,(\ref{Constraint}) from the right by the inverse matrix $\mathbb{S}^{-1}(\Omega,x)$ and then contracting both sides of the equation by ${\rm d}x^\mu$, and recalling that the Fock-Ivanenko matrix-valued line element ${\rm d}s=\gamma_\mu{\rm d}x^\mu$ \cite{FockIvanenko2} is written as ${\rm d}\tilde{s}\equiv\tilde{\gamma}_\mu{\rm d}x^\mu$ in the conformal frame, Eq.\,(\ref{Constraint}) takes the form
\begin{equation}\label{FockIvanenkoConstraint}
    \mathbb{S}\tilde{D}\,\mathbb{S}^{-1}=\frac{\Omega_{,\mu}\tilde{\gamma}^\mu}{2\Omega}{\rm d}\tilde{s},
\end{equation}
where $\tilde{D}$ stands for the covariant differentiation operator in the conformal frame. This is a total differential equation that remarkably involves the Fock-Ivanenko on the right-hand side in a fundamental way. 

In order to find the explicit expression of the functional $\mathbb{S}(\Omega,x)$, we solve the first-order differential equation on the left in Eq.\,(\ref{2OtherConstraints}). For that purpose, we choose the following general ansatz for $\mathbb{S}(\Omega,x)$,
\begin{equation}\label{Generalansatz}
\mathbb{S}(\Omega,x)=\Omega_{,\mu}\gamma^\mu f(\Omega,x)+\varepsilon^{\mu\nu\rho\sigma}\Omega_{,\mu}\gamma_\nu\gamma_\rho\gamma_\sigma\, h(\Omega,x),    
\end{equation}
where $f(\Omega,x)$ and $h(\Omega,x)$ are each an arbitrary complex scalar functional of $\Omega$, and $\varepsilon^{\mu\nu\rho\sigma}$ is the totally antisymmetric Levi-Civita tensor. Indeed, the combination in expression (\ref{Generalansatz}) is the only way of building an invariant matrix-valued functional out of the gamma matrices by contracting the indices of the latter with the only available vector $\Omega_{,\mu}$. Inserting the ansatz (\ref{Generalansatz}) into the first equation in (\ref{2OtherConstraints}) and separating those independent terms made of the independent elements of the Dirac algebra $\gamma^\mu$ and $\varepsilon^{\mu\nu\rho\sigma}\gamma_\nu\gamma_\rho\gamma_\sigma$, we arrive at the following two independent differential equations for $f(\Omega,x)$ and $h(\Omega,x)$:  
\begin{align}\label{DifferentialEquations}
\Omega\Omega^{,\mu}f_{,\mu}+\Omega\,\Box\Omega f-\Omega_{,\mu}\Omega^{,\mu}f&=0,\\
\varepsilon^{\lambda\nu\rho\sigma}\gamma^\mu\gamma_\nu\gamma_\rho\gamma_\sigma\left(\Omega\Omega_{,\lambda}h_{,\mu}+\Omega\Omega_{,\lambda\mu}h-\Omega_{,\lambda}\Omega_{,\mu}h\right)&=0.
\end{align}
Dividing both equations by $\Omega^2$ and then rearranging their terms, the two equations take the following forms, respectively:
\begin{equation}\label{SimplifiedDifferentialEquations}
\nabla^\mu\left(\frac{\Omega_{,\mu}}{\Omega}f\right)=0,\qquad
\varepsilon^{\mu\nu\rho\sigma}\gamma_\mu\gamma_\nu\gamma_\rho\gamma_\sigma\nabla^
\lambda\left(\frac{\Omega_{,\lambda}}{\Omega}h\right)=0.
\end{equation}
Therefore, we conclude that $f(\Omega,x)$ and $h(\Omega,x)$ are functionals that render $\Omega_{,\mu}f/\Omega$ and $\Omega_{,\mu}h/\Omega$ divergence-free vector fields, \textit{i.e}, conserved vector fields in the spacetime region where the conformal function $\Omega(x)$ is defined. 

Next, the conformal transformation $\psi=\Omega^{\frac{3}{2}}\mathbb{S}(\Omega,x)\tilde{\psi}$ required to guarantee the conformal invariance of Eq.\,(\ref{ModifiedSD}) should also preserve the structure and the usual probability amplitude interpretation assigned to a spinor. This can be achieved by starting from the following transformation of the integrated probability density over an arbitrary spatial volume $V$:
\begin{equation}
  \int\psi^\dagger\psi\,{\rm d}V=\int\tilde{\psi}^\dagger\,\mathbb{S}^\dagger\mathbb{S}\,\tilde{\psi}\,{\rm d}\tilde{V}.  
\end{equation}
In writing the right-hand side of this equality, we used the fact that $\Omega^3{\rm d}V={\rm d}\tilde{V}$. For unitarity to be preserved under such a conformal transformation, we only need then to impose the extra condition that $\mathbb{S}^\dagger\mathbb{S}$ be the identity matrix. Using Eq.\,(\ref{Generalansatz}) and the identity $\varepsilon^{\mu\nu\rho\sigma}\gamma_{\nu}\gamma_{\rho}\gamma_{\sigma}=-6i\gamma^\mu\gamma^5$, where $\gamma^5=i\gamma^0\gamma^1\gamma^2\gamma^3$, we translate this unitarity condition into
\begin{equation}\label{ExtraCondition}
\Omega_{,\mu}\Omega_{,\nu}\gamma^{\mu\dagger}\gamma^\nu\left[|f|^2+36|h|^2+6i\left(f^*h-h^*f\right)\gamma^5\right]=\mathbb{1}.
\end{equation}
Thus, whereas equations (\ref{SimplifiedDifferentialEquations}) give the functionals $f(\Omega,x)$ and $h(\Omega,x)$ only up to an arbitrary multiplicative complex constant for each, the extra condition (\ref{ExtraCondition}) is what fixes those arbitrary two complex (four real) constants. Of course, the existence of those two functionals $f(\Omega,x)$ and $h(\Omega,x)$ obeying simultaneously equations (\ref{SimplifiedDifferentialEquations}) and (\ref{ExtraCondition}) should be examined case by case. In other words, a given conformal factor $\Omega(x)$ may or may not lead to actual functionals $f(\Omega,x)$ and $h(\Omega,x)$. This is in contrast to the conformal transformation $\psi=\Omega^{\frac{3}{2}}\tilde{\psi}$ required by the Dirac equation.

\section{Revisiting the Schr\"odinger-Dirac equation in the presence of the Maxwell field}\label{sec:SchroDiMaxwell}

We examine here the Schr\"odinger-Dirac equation coupled to the Maxwell field $A_\mu$. For that purpose, we start by deriving the equation for the case where the particle is coupled to the Maxwell field only through the usual minimal-coupling prescription. In other words, one introduces the Maxwell field into the Dirac equation only by replacing the spin-covariant derivative operator $D_\mu$ by $D_\mu+ieA_\mu$ inside the Dirac equation. 

Just as in the derivation we displayed in Appendix \ref{sec:AppDerivation} for the Schr\"odinger-Dirac equation of neutral spinors, we may also now apply simply the derivative operator $\gamma^\nu D_\nu$ from the left to the Dirac equation minimally coupled to the Maxwell field. This yields,
\begin{align}\label{SchroDiracMaxwell}
    \gamma^\nu D_\nu(i\gamma^\mu D_\mu-e\gamma^\mu A_\mu-m)\,\psi&=0\nonumber\\
    \Longrightarrow\;\left(\!\,^A\mathcal{D}_\mu \,\!^A\mathcal{D}^\mu+2ie\Sigma^{\mu\nu}F_{\mu\nu}+\tfrac{1}{4}R+m^2\right)\psi&=0.
\end{align}
We introduced here the gauge- and spin-covariant derivative operator $\!\,^A\mathcal{D}_{\mu}=\mathcal{D}_\mu+ieA_\mu$, as well as the field strength tensor $F_{\mu\nu}=\partial_\mu A_\nu-\partial_\nu A_\mu$. The gauge invariance of the resulting equation is guaranteed by the gauge invariance of the Dirac equation itself, as can easily be seen from the first line. The term $2ie\Sigma^{\mu\nu}F_{\mu\nu}$ gives rise to the familiar Pauli term in the nonrelativistic Pauli-Schr\"odinger equation \cite{Pauli1}. The usual $g=2$ factor of the electron is indeed easily recognized by writing that term with the help of the tensor $\sigma_{\mu\nu}=\frac{i}{2}[\gamma_\mu,\gamma_\nu]$, for then $2ie\Sigma^{\mu\nu}F_{\mu\nu}$ takes the more familiar textbook form $\frac{e}{2}\sigma_{\mu\nu}F^{\mu\nu}$ \cite{QFT2}. In fact, if $F_{\mu\nu}$ represents a magnetic field $\bf{B}$ lying along the $z$ direction in a given reference frame in four-dimensional spacetime, then
\begin{equation}\label{WithB-Field}
\frac{e}{2}\sigma_{\mu\nu}F^{\mu\nu}=-e
\begin{pmatrix}
{\bf{s}}.{\bf{B}} & 0\\
0  & {\bf{s}}.{\bf{B}}
\end{pmatrix}.
\end{equation}
The three-dimensional vector $\bf s$ is the spin vector $\frac{1}{2}\boldsymbol\sigma$, where the $2\times2$ matrices $\boldsymbol\sigma$ are the Pauli matrices. Note that the same equation (\ref{SchroDiracMaxwell}) is also what one obtains when one applies either $i\gamma^\mu D_\mu-e\gamma^\mu A_\mu$ or $i\gamma^\mu D_\mu-e\gamma^\mu A_\mu+m$ to the Dirac equation minimally coupled the the Maxwell field.
\subsection{Particles with an anomalous magnetic moment}
It is well known that for particles with an anomalous magnetic moment (like the electron, the proton, the neutron, the muon, and so on) one may write a Dirac equation for such particles as follows \cite{Pauli2} (see also, \cite{Foldy,Behncke}),
\begin{align}\label{AnomalousDirac}
    \left(i\gamma^\mu D_\mu-e\gamma^\mu A_\mu-\tfrac{i}{4}\mu_0 \gamma^\mu\gamma^\nu F_{\mu\nu}-m\right)\,\psi&=0.
\end{align}
We introduced the factor $\frac{1}{4}$ multiplying the constant  $\mu_0$ (as done in Ref.\,\cite{Behncke}) in order to simplify our subsequent equations. As we shall show shortly, the constant $\mu_0$ is what would give rise the anomalous magnetic moment of the particle in the curved spacetime. First, note that Eq.\,(\ref{AnomalousDirac}) is clearly gauge invariant. Therefore, we may legitimately apply a linear differential operator to this equation from the left to obtain a second-order differential equation for particles with an anomalous magnetic moment that would also be gauge invariant. For that purpose, we apply to Eq.\,(\ref{AnomalousDirac}) from the left the differential operator $\gamma^\nu D_\nu$. This yields,
\begin{multline}\label{SchroDiracAnomalous}
    \left(\,\!^A\mathcal{D}_\mu \!\,^A\mathcal{D}^\mu+\mu_0F^{\mu\nu}\gamma_\mu\!\,^A\!D_\nu+2i[e+\mu_0 m]\Sigma^{\mu\nu}F_{\mu\nu}+\tfrac{1}{4}R+m^2\right.\\ 
\left.-\mu_0^2\,\Sigma^{\mu\nu}\Sigma^{\rho\lambda}F_{\mu\nu}F_{\rho\lambda}-2\pi\mu_0j_\mu\gamma^\mu\right)\psi=0.
\end{multline}
We introduced here the charged current $j^\mu$ to which the Maxwell field couples thanks to Maxwell's second equation $\nabla_\mu F^{\mu\nu}=4\pi j^\nu$. Also, we set $\,^A\!D_\mu =D_\mu+ieA_\mu$.

From Eq.\,(\ref{SchroDiracAnomalous}), we start to see a gradual emergence of a general pattern each time one extracts a second-order differential equation from the Dirac equation. In fact, when starting from the Dirac equation minimally coupled to the Maxwell field, the second-order differential equation one arrives at displays, as we saw below Eq.\,(\ref{SchroDiracMaxwell}), a coupling of the particle to the magnetic field though the magnetic moment of the former. The usual $g$-factor for the electron predicted by the Dirac equation emerges from the resulting term $2ie\Sigma^{\mu\nu}F_{\mu\nu}$. What we get from Eq.\,(\ref{SchroDiracAnomalous}) is a similar term, but in which the electric charge $e$ is augmented by $\mu_0 m$. This shows that the anomalous magnetic moment of the particle in this case comes from $\mu_0$, and the effective $g$-factor of the particle reads $2(1+\frac{\mu_0}{e}m)$.

However, in addition to this coupling, the other manifestation of the anomalous magnetic moment of the particle consists of inducing a coupling of the gauge-invariant gradient $\!\,^A\!D_\mu\psi$ of the spinor field with the field strength $F_{\mu\nu}$ as given by the second term in Eq.\,(\ref{SchroDiracAnomalous}), as well as a direct coupling of the spinor field $\psi$ with the source current $j_\mu$ as given by the last term in Eq.\,(\ref{SchroDiracAnomalous}). Furthermore, the anomalous magnetic moment also induces a direct coupling of the spinor field with the field strength ---\,as given by the term before last in Eq.\,(\ref{SchroDiracAnomalous})\,--- that is second-order in the spin. In the next subsection, we investigate how this pattern generalizes further in the case of particles with higher electric and magnetic moments. 

It is noteworthy to point out here that if no charged current  couples to the Maxwell field, other than the one coming from the Dirac current $j^\nu=e\bar\psi\gamma^\nu\psi$, then Eq.\,(\ref{SchroDiracAnomalous}) takes again a nonlinear form as does Eq.\,(\ref{SchroDiracNonLinear}).
\subsection{Particles with higher electric and magnetic moments}
It was shown by Foldy in Ref.\,\cite{Foldy} that the most general Lorentz invariant, local and gauge invariant Dirac equation that is linear both in the spinor field $\psi$ and in the external electromagnetic field $A_a$, has the following form in Minkowski spacetime:
\begin{align}\label{Foldy}
    \left(i\gamma^a \partial_a-\sum_{n=0}^\infty e_n\Box^n \gamma^a A_{a}-\tfrac{i}{4}\sum_{n=0}^\infty \mu_n\Box^n \gamma^a\gamma^b F_{ab}-m\right)\,\psi&=0.
\end{align}
In Ref.\,\cite{Foldy}, the operator $\Box$ stands for the flat-spacetime d'Alembertian operator $\eta^{ab}\partial_a\partial_b$. Comparing this equation with Eq.\,(\ref{AnomalousDirac}), one learns that $e_0$ should be identified with the electric charge of the particle and $\mu_0$ should be identified with the anomalous magnetic moment of the particle. The additional terms with $n\geq1$ represent a convenient approximation for what a complete field theoretical treatment yields for the interaction of the Dirac particle with the external electromagnetic field created by a certain charge and current distribution \cite{Foldy}. In Ref.\,\cite{Behncke}, Behncke suggested that those additional terms  might also be viewed as describing the interaction of an extended charge. 

The way one arrives at all the additional terms appearing in Eq.\,(\ref{Foldy}) in flat spacetime is to consider all the possible contractions $\gamma^a...\gamma^b\gamma^c\partial_a...\partial_b A_c$. Then, working within the Lorenz gauge $\partial_a A^a=0$, one is able to show \cite{Foldy} that such combinations split only into two different classes of terms. One class consists of terms of the form $\Box^n\gamma^aA_a$, and another class consists of terms of the form $\Box^n\gamma^a\gamma^bF_{ab}$. Therefore, after using Maxwell's equation in the Lorenz gauge, $\Box A_a=4\pi j_a$, all the additional terms in Eq.\,(\ref{Foldy}) can be written in terms of the charged current $j_\mu$ and its derivatives: $e_n\Box^{n-1}\gamma^a j_a$ for the first class, and $\mu_n\Box^{n-1}\gamma^a\gamma^b(\partial_a j_b-\partial_b j_a)$ for the second class \cite{Foldy}. 

In curved spacetime, one is not only forbidden to use the non-covariant terms $\partial_\mu...\partial_\nu A_\rho$, but even Maxwell's equation takes the more involved form $\Box A_\mu-R_\mu\,\!^{\nu}A_\nu-\nabla_\mu\nabla^\nu A_\nu=4\pi j_\mu$. Covariant derivatives of $A_\mu$ should thus be used instead, and then all possible gauge-invariant contractions of the latter with the gamma matrices should be considered. Fortunately, however, thanks to the gauge-symmetry invariance of the current $j_\mu$ we may easily adopt Foldy's final form of the equation to curved spacetimes. Indeed, the general covariant form of Eq.\,(\ref{Foldy}) in terms of the gauge invariant vector $j_\mu$ and the Maxwell tensor $F_{\mu\nu}$ will then take the form
\begin{equation}\label{GeneralizedAnomalousDirac}
    \left(i\gamma^\mu D_\mu\!-\!e\gamma^\mu A_{\mu}\!-\!\sum_{n=0}^\infty e_n\Box^{n}\gamma^\mu j_{\mu}\!-\!\tfrac{i}{4}\sum_{n=0}^\infty \mu_n\Box^n \gamma^\mu\gamma^\nu F_{\mu\nu}\!-\!m\right)\!\,\psi=0.
\end{equation}
We have denoted here the electric charge of the particle by $e$, instead of $e_0$, in order to make the infinite series in $\Box^n\gamma^\mu j_\mu$ start from $n=0$ as does the one involving the field strength $F_{\mu\nu}$. Thus, the constant $e_0$ stands here for the lowest-order electric moment of the particle. On the other hand, the d'Alembertian $\Box$ in our equation (\ref{GeneralizedAnomalousDirac}) is now the generally covariant one, $g^{\mu\nu}\mathcal{D}_\mu\mathcal{D}_\nu$. Applying the operator $\gamma^\nu D_\nu$ from the left to Eq.\,(\ref{GeneralizedAnomalousDirac}), yields
\begin{multline}\label{GeneralizedSchroDiracAnomalous}
    \Bigg[\!\,^A\mathcal{D}_\mu \,\!^A\mathcal{D}^\mu+\sum_{n=0}^\infty\left(2ie_n\Box^nj^\nu+\mu_n\Box^nF^{\mu\nu}\gamma_\mu\right)\!\,^A\!D_\nu\\+2i\Big(e+m\sum_{n=0}^\infty\mu_n\Box^n \Big)\Sigma^{\mu\nu}F_{\mu\nu}+\tfrac{1}{4}R+m^2 -\sum_{n,m=0}^\infty e_ne_m(\Box^nj_\mu)(\Box^mj^\mu)\\
    -\Sigma^{\mu\nu}\Sigma^{\rho\lambda}\sum_{n,m=0}^\infty\mu_n\mu_m(\Box^nF_{\mu\nu})(\Box^mF_{\rho\lambda})+4i\Sigma^{\mu\nu}\sum_{n=0}^\infty e_n\nabla_\mu\Box^nj_\nu \\
    +i\sum_{n=0}^\infty e_n\nabla_\mu\Box^nj^\mu-\gamma^\rho\Sigma^{\mu\nu}\sum_{n=0}^\infty\mu_n\nabla_\rho\Box^nF_{\mu\nu}
+i\gamma^\mu\sum_{n,m=0}^\infty\mu_ne_m\left(\Box^nF_{\mu\nu}\right)\left(\Box^mj^\nu\right) \Bigg]\psi=0.
\end{multline}
Several interesting observations can be made by inspecting this equation. The first thing we notice in this equation when comparing it with Eq.\,(\ref{SchroDiracAnomalous}) is that the anomalous magnetic moment of the particle is displayed now as an infinite-series correction to the standard magnetic moment one derives from the Dirac equation. In addition, however, a similar spin interaction emerges now thanks to the coupling of the spin with the higher-derivatives $\nabla_\mu\Box^nj_\nu$ of the current, as well as a coupling of the spin with the higher-derivatives $\nabla_\rho\Box^nF_{\mu\nu}$ of the field strength. The next observation that can be made is that besides the coupling of the gauge-invariant gradient $\,\!^A\!D_\mu\psi$ of the spinor field with the higher derivatives $\Box^nF_{\mu\nu}$ of field strength, there is also a directed coupling of that gradient with the higher derivatives $\Box^nj_\mu$ of the current. On the other hand, the spin-spin interaction gets now generalized into an interaction that includes all possible pair products of higher-derivative terms $\Box^nF_{\mu\nu}$ of the field strength. We also notice the emergence now of an interaction that is quadratic in the higher-derivative terms $\Box^nj_\mu$ that couple directly to the spinor field. Furthermore, the last term also involves a direct coupling of the spinor field with a product of cross terms made of higher derivatives of both the field strength and the current. The divergence terms $\nabla_\mu\Box^nj^\mu$ did not appear before taking into account the higher moments of the particle because at the lowest order we considered previously, only the vanishing term $\nabla_\mu j^\mu$ would have been included.

The last observation we would like to make here is the remarkable appearance in Eq.\,(\ref{GeneralizedSchroDiracAnomalous}) of terms very similar to those that emerge in the form of curvature terms from the study of the expectation value of the stress-energy tensor in curved spacetime \cite{QFT1,QFT2}. In fact, moving the covariant derivative $\nabla_\mu$ past the d'Alembertian $\Box$ in the non-vanishing lowest-order term $e_1\nabla_\mu\Box j^\mu$, as well as in the lowest-order term in the spin, $4ie_1\Sigma^{\mu\nu}\nabla_\mu\Box j_\nu$, we learn that these two terms read
\begin{align}\label{manipulation1}
    ie_1\nabla_\mu\Box j^\mu&=-ie_1\left(\frac{1}{2}j^\mu R_{,\mu} +3\nabla^\mu j^\nu R_{\mu\nu}\right),\\\label{manipulation2}
    4ie_1\Sigma^{\mu\nu}\nabla_\mu\Box j_\nu&=4ie_1\Sigma^{\mu\nu}\left(\Box\nabla_\mu j_\nu+j^\rho\nabla_\nu R_{\mu\rho}-2\nabla^\rho j_\lambda R_{\mu\rho\nu}\,^\lambda-\nabla^\rho j_\nu R_{\mu\rho}\right).
\end{align}
In deriving the first identity we also used the conservation equation $\nabla_\mu j^\mu=0$, as well as the contracted Bianchi identity $\nabla_\mu R^\mu\,_\nu=\frac{1}{2} R_{,\nu}$. On the other hand, recalling that the second Maxwell equation can also be written in terms of the field strength tensor as
\begin{equation}\label{Maxwell}
\Box F_{\mu\nu}+2R_{\mu\rho\nu\lambda}F^{\rho\lambda}+R_{\mu\rho}F^{\rho}\,_\nu-R_{\mu\rho}F^{\rho}\,_\nu=4\pi \left(\nabla_\mu j_{\nu}-\nabla_\nu j_{\mu}\right),    
\end{equation} 
the contraction of both sides of the latter equation by $\Sigma^{\mu\nu}$ turns Eq.\,(\ref{manipulation2}) into
\begin{multline}\label{manipulation3}
    \!\!\!\!\!4ie_1\Sigma^{\mu\nu}\nabla_\mu\Box j_\nu=4ie_1\Sigma^{\mu\nu}\left(\frac{1}{8\pi}\Box^2F_{\mu\nu}+\frac{1}{4\pi}\Box R_{\mu\rho\nu\lambda} F^{\rho\lambda}+\frac{1}{4\pi}R_{\mu\rho\nu\lambda}\Box F^{\rho\lambda}+\frac{1}{2\pi}\nabla_\delta R_{\mu\rho\nu\lambda}\nabla^\delta F^{\rho\lambda}\right.\\
    \left.+\frac{1}{4\pi}\Box R_{\mu\rho} F^{\rho}\,_\nu+\frac{1}{4\pi}R_{\mu\rho} \Box F^{\rho}\,_\nu+\frac{1}{2\pi}\nabla_\lambda R_{\mu\rho}\nabla^\lambda F^{\rho}\,_\nu+j^\rho\nabla_\nu R_{\mu\rho}-2\nabla^\rho j_\lambda R_{\mu\rho\nu}\,^\lambda-\nabla^\rho j_\nu R_{\mu\rho}\vphantom{\frac{1}{8\pi}}\right).
\end{multline}
From Eq.\,(\ref{manipulation1}), we learn that the lowest-order term $ie_1\nabla_\mu\Box j^\mu$ brings a correction to the term $-2\pi\mu_0j_\mu\gamma^\mu$ of Eq.\,(\ref{SchroDiracAnomalous}) that is equal to $-\frac{i}{2}e_1 j_\mu R^{,\mu}$, and brings into Eq.\,(\ref{SchroDiracAnomalous}) an additional direct coupling between the spinor field and the gradient of the current that is equal to $-3ie_1\nabla^\mu j_\nu R_{\mu\nu}$. Inspection of Eq.\,(\ref{manipulation3}), on the other hand, reveals that the lowest-order term in the particle's spin, $4ie_1\Sigma^{\mu\nu}\nabla_\mu\Box j_\nu$, contributes with various correcting terms to the anomalous magnetic moment of the particle. Those corrections include terms that are proportional to $e_1/4\pi$, to the Riemann tensor $R_{\mu\nu\rho\lambda}$, to the Ricci tensor $R_{\mu\nu}$ and to the second-order derivatives of these tensors, $\Box R_{\mu\nu\rho\lambda}$, $\Box R_{\mu\nu}$, etc. Furthermore, iterating similar manipulations as those leading to Eqs.\,(\ref{manipulation1}) and (\ref{manipulation3}) shows that the higher-derivative terms $ie_n\nabla_\mu\Box^n j^\mu$ and $4i\Sigma^{\mu\nu}e_n\nabla_\mu\Box^n j_\nu$ can also be decomposed into correcting terms for the anomalous magnetic moment in Eq.\,(\ref{GeneralizedSchroDiracAnomalous}). Those terms contain not only higher derivatives of the Riemann and Ricci tensors, but also products of the latter, like $R_{\mu\lambda}R^\lambda\,_{\nu}$, $R_{\rho\lambda}R^\lambda\,_{\alpha\mu\nu}$, $R_{\mu\nu\rho\lambda}R^\lambda\,_{\alpha\beta\delta}$, etc. 

The above observations become particularly interesting when recalling, as already mentioned below Eq.\,(\ref{Foldy}), that the additional terms brought to the Dirac equation are supposed to represent approximations of higher-order interaction terms that emerge from a more complete quantum field theoretical treatment. Therefore, it is not surprising that one recovers also corrections involving curvature terms and their derivatives and contractions just as one does within curved-spacetime quantum field theory. In contrast to the latter, however, the curvature terms that arose here have nothing to do with quantum vacuum expectation values. Those terms appeared here simply as gravity-induced correction terms inside the second-order differential Schr\"odinger-Dirac equation that governs the dynamics of a Dirac particle inside an external electromagnetic field within general relativity.
\section{Revisiting Lichnerowicz's vanishing theorem}\label{sec:Lichnerowicz}
As it was discussed in Sec.\,\ref{sec:Intro}, the identity $\slashed{D}^2\!=\!g^{\mu\nu}\mathcal{D}_\mu \mathcal{D}_\nu+\frac{1}{4}R$ behind the derivation of the Schr\"odinger-Dirac equation (\ref{SchroDirac}) was re-derived independently by Lichnerowicz in Ref.\,\cite{Lichnerowicz}. The identity was used by Lichnerowicz to derive an important theorem for Riemannian spin manifolds. The theorem, sometimes called the vanishing theorem \cite{SpinGeometry}, asserts that a compact spin manifold of positive scalar curvature does not admit any non-zero harmonic spinor. Conversely, the theorem implies that a compact spin manifold with an identically vanishing scalar curvature can only admit parallel spinors. Given the results of Sec.\,\ref{sec:ConfSchroDirac} concerning the Schr\"odinger-Dirac equation in the conformal frame, it is of great interest to also examine here the fate of the vanishing theorem in light of those results. Let us first outline the usual proof of the vanishing theorem.

We call a harmonic spinor $\psi$ any spinor satisfying the equation $\slashed{D}^2\psi=0$ on a compact spin manifold, \textit{i.e.}, when $\slashed{D}\psi=0$. We say that a Riemannian spacetime is a compact spin manifold of positive scalar curvature if the spacetime is an orientable compact Riemannian manifold with a spin structure on its tangent bundle, and if its Ricci scalar $R$ is everywhere positive or zero ($R\geq 0$) without being identically zero \cite{SpinGeometry}. Therefore, multiplying the Lichnerowicz-Schr\"odinger identity (which reads $\slashed{D}^2=-g^{\mu\nu}\mathcal{D}_\mu \mathcal{D}_\nu+\frac{1}{4}R$ in the spin-manifold thanks to the anti-commutation relations in Riemannian spacetimes $\{\gamma_\mu,\gamma_\nu\}=-2g_{\mu\nu}$ \cite{Lichnerowicz}) by $\psi^\dagger$ from the left and by $\psi$ from the right, and then integrating by parts both sides of the equation over the four-volume of the compact manifold leads to the following identity:
\begin{align}\label{Theorem}
    \!\int\!{\rm d}^4x\sqrt{g}\,\psi^\dagger\slashed{D}^2\psi&=-\!\int\!{\rm d}^4x\!\sqrt{g}\,\psi^\dagger\mathcal{D}^\mu \mathcal{D}_\mu\psi+\tfrac{1}{4}\!\int\!{\rm d}^4x\!\sqrt{g}\,\psi^\dagger R\,\psi\nonumber\\
    &=\int{\rm d}^4x\sqrt{g}\,|D\;\!\psi|^2+\tfrac{1}{4}\int{\rm d}^4x\sqrt{g}\, R\,|\psi|^2.
\end{align}
Here, we denoted by $g$ the positive determinant of the Riemannian metric. In the mathematics literature, it is sometimes this identity which is called a theorem while the consequences derived from it are called corollaries \cite{SpinGeometry}\footnote{Other names are also found in the mathematics literature. For example, the manipulation (\ref{Theorem}) is sometimes called the Bochner technique, whereas the Lichnerowicz-Schr\"odinger identity and its variants are called the Weitzenb\"ock formulas \cite{Petersen}, the Bochner-Weitzenb\"ock formulas \cite{Li}, or simply the Weitzenb\"ock decomposition \cite{Besse}.}. In fact, we clearly see that if $R\geq0$ the right-hand side of Eq.\,(\ref{Theorem}) cannot vanish unless $\psi=0$, which means that no nonzero harmonic spinor can exist. Conversely, if $R$ vanishes everywhere, then only harmonic spinors such that $|D\,\psi|^2=0$ can exist in the manifold, which means that only parallel spinors are admitted in such a manifold. It follows then that if ever a nonzero harmonic spinor exists in a manifold of non-negative curvature, that harmonic spinor is parallel and the Ricci scalar $R$ of the manifold should vanish everywhere.

When going to the conformal frame, it is clear that by working right from the start with the metric $\tilde{g}_{\mu\nu}$ and the spinor $\tilde{\psi}$, all the steps leading to identity (\ref{Theorem}) remain unaltered. All the terms of the identity simply become decorated all over the place by tildes. This implies then that the corollaries derived in the previous paragraph from identity (\ref{Theorem}) remain true for the manifold obtained by conformally deforming the original spin manifold. Of course, it goes without saying that a positive-curvature manifold does not conformally transform necessarily into a positive-curvature manifold. This can be seen from the conformal transformation of the Ricci scalar $R$ as given by the last identity in Eq.\,(\ref{ConformalConnections}). But the point here is that once such a conformal manifold is obtained, this first method suggests that one reaches the same conclusions about the relation between the sign of the scalar curvature $\tilde{R}$ of the new manifold and the existence of harmonic spinors $\tilde{\psi}$ in the manifold. 

However, if we first start from identity (\ref{Theorem}) which holds in the original spin manifold, then, as we show in detail in Appendix \ref{sec:AppConfLichSchro}, the conformally transformed version of identity (\ref{Theorem}) takes the following form instead:
\begin{align}\label{ConfTheorem}
    \!\int{\rm d}^4x\!\sqrt{\tilde{g}}\,\Omega\,\tilde{\psi}^\dagger\tilde{\slashed{D}}^2\tilde{\psi}&=\!\int\!{\rm d}^4x\sqrt{\tilde{g}}\,\Omega\,|\tilde{D}\,\tilde{\psi}|^2+\tfrac{1}{4}\!\int\!{\rm d}^4x\sqrt{\tilde{g}}\,\Omega\,\tilde{R}\,|\tilde{\psi}|^2+\int{\rm d}^4x\sqrt{\tilde{g}}\,\Omega^{,\mu}\tilde{\psi}^\dagger\tilde{D}_\mu\tilde{\psi}.
\end{align}
It is clear from this result that the corollaries derived from identity (\ref{Theorem}) do not hold anymore in the conformal manifold. In fact, according to identity (\ref{ConfTheorem}) a spin manifold of strictly positive scalar curvature $\tilde{R}$ does not imply that nonzero harmonic spinors cannot exist, and neither does the condition $\tilde{R}=0$ imply that only harmonic spinors such that $|\tilde{D}\,\tilde{\psi}|^2=0$ can exist in the manifold. This is due to the last term which consists of an integral that could be positive, zero or negative, depending on the conformal function $\Omega(x)$ that couples via its gradient to the term $\tilde{\psi}^\dagger\tilde{D}_\mu\tilde{\psi}$. See Refs.\,\cite{Hijazi,Axioms} for other results on harmonic functions in conformal spin manifolds.
\section{Summary and conclusion}\label{sec:Summary}
We have revisited the usual derivation of the Schr\"odinger-Dirac equation and shown that depending on whether one initially assumes the Dirac equation to hold or not, there are two possible forms for the resulting second-order differential equation for the spinor. The first case leads to the usual Schr\"odinger-Dirac equation, whereas in the second case a slightly different second-order differential equation is obtained. We then showed that a nonlinear covariant Gross-Pitaevskii-like equation emerges from the Schr\"odinger-Dirac equation when one takes into account the fact that the spinor field self-interacts via its coupling to the curvature of spacetime. We extracted form the equation the gravitationally induced analog for fermions of the scattering length of bosons. Such a length, given by $g=4\pi\hbar^2 a_s/m$ for bosons, is found here to take on the much smaller value $mG/4c^2$ for fermions coupled to gravity. 

Note that one of the possible applications of such a nonlinear equation would be in the study of matter under gravitational collapse. In fact, it is well known that besides the traditional methods relying on the perfect-fluid approximation or the equation of state of matter, one can study the equilibrium configurations of a self-gravitating system of fermions by relying on the Dirac equation in curved spacetime coupled to the Einstein equations \cite{RuffiniBonazzola}. In the latter approach, one inserts into the right-hand side of Einstein's equations the expectation value of the energy-momentum tensor of the fermions. In contrast, Eq.\,(\ref{SchroDiracNonLinear}) offers us the possibility of tackling the problem of the equation of state of dense compact stars in a more closed form. Solving Eq.\,(\ref{SchroDiracNonLinear}) for such a system should indeed lead to a more accurate dynamics as the nonlinearity of the wave equation takes into account the gravitational back-reaction of the fermion fluid. Examining closer this possible application will be the subject of a future work.

The conformal noninvariance of the Schr\"odinger-Dirac equation is also examined in detail. This allowed us to build a modified Schr\"odinger-Dirac equation that is conformally invariant in analogy to the well-known non-minimally coupled scalar field in a generalized Klein-Gordon equation. However, the required conformal transformation of the spinor for such a modified Schr\"odinger-Dirac equation is found to require a conformal factor that is a matrix-valued functional of the conformal factor $\Omega(x)$. This matrix-valued functional is found to obey a first-order differential equation that involves the Fock-Ivanenko line element in a fundamental way. A supplementary condition on the conformal factor, as dictated by the conservation of unitarity under the conformal transformation, is provided.

The coupling of the Schr\"odinger-Dirac equation to the Maxwell field is then revisited and the detailed derivation giving rise to the emergence of the correct gyromagnetic moment of the particle is recalled. Using the same procedure, and starting form the generalized Dirac equation proposed by Pauli for particles with an anomalous magnetic moment, we derived a new Schr\"odinger-Dirac equation for such particles. We showed that just as the Schr\"odinger-Dirac equation gives rise to the exact $g$-factor from the Dirac equation, the new Schr\"odinger-Dirac equation gives rise to a correction for the $g$-factor as well as additional terms that represent new interactions of the particle with the external electromagnetic field and its source. Those results allowed us to generalize even further our equation by appealing to Foldy's equation for describing particles with higher electric and magnetic moments. Although being much more involved, the resulting equation displayed a remarkably simple pattern. The latter consists of higher-order terms representing corrections to the electric and magnetic moments that involve spacetime curvature terms and their derivatives in analogy to what is found within quantum field theory when computing the expectation value of the stress-energy tensor in curved spacetime. We believe that this result could bring new applications of the generalized Schr\"odinger-Dirac equation and new insights on the latter in future work.

Finally, we revisited Lichnerowicz's vanishing theorem under the conformal transformation. We showed that the conformal noninvariance of the identity behind the theorem leads to interesting implications. We showed that the identity takes on different forms in the conformal frame depending on how one extracts the transformed identity. If one starts by considering the conformal frame and its conformal spinor, then the same steps leading to the derivation of the identity behind the theorem in the original frame also give rise to the same identity in the conformal frame. This implies that the corollaries one extracts for conformal harmonic spinors and a compact manifold in the conformal frame are identical to the ones extracted for those entities in the original frame. If, on the other hand, one starts by transforming both sides of the identity that holds in the original frame, then one does not recover the same form of the identity in the new frame. This implies that the corollaries one extracts in the new frame from the resulting identity do not hold anymore in the latter frame. This comes about because in the second method one has to integrate by parts.

\section*{Addendum}
After this paper has been published, we learned about Addendum \cite{KayNote} to Ref.\,\cite{SchrodingerNote}, where the authors pointed out that Asher Peres has also rediscovered in Refs.\,\cite{Peres1,Peres2} the square of the curved-spacetime Dirac operator that we called here the Lichnerowicz-Schr\"odinger identity. One of the authors of the present paper (FH) is grateful to Prof. Bernard S. Kay for having communicated to him this information.

\section*{Acknowledgments}
The authors are grateful to the anonymous referees for the very helpful and constructive comments. This work was supported by the Natural Sciences and Engineering Research Council of Canada (NSERC) Discovery Grant No. RGPIN-2017-05388; and by the Fonds de Recherche du Québec - Nature et Technologies (FRQNT). PS acknowledges support from Bishop's University via the Graduate Entrance Scholarship award and from Bishop's University Research Assistantship award.

\appendix
\section{Derivation of the Sch\"odinger-Dirac equation} \label{sec:AppDerivation} 
The derivation of the Schr\"odinger-Dirac equation (\ref{SchroDirac}) relies on the Lichnerowicz-Schr\"odinger identity $\slashed{D}^2=g^{\mu\nu}\mathcal{D}_\mu \mathcal{D}_\nu+\frac{1}{4}R$. The proof of the Lichnerowicz-Schr\"odinger identity starts by recalling that the vierbeins are covariantly constant: $\mathcal{D}_\mu e^\nu_a=0$. This implies that $\mathcal{D}_\mu \gamma^\nu=0$, where the action of the total derivative $\mathcal{D}_\mu$ on the matrix-valued tensor $\gamma^\nu$ is defined by $\mathcal{D}_\mu\gamma^\nu=\nabla_\mu\gamma^\nu+[\Lambda_\mu,\gamma^\nu]$. Therefore, we have $\gamma^\mu D_\mu\left(\gamma^\nu D_\nu\right)=\gamma^\mu \gamma^\nu\mathcal{D}_\mu\mathcal{D}_\nu$. The remainder of the proof then proceeds as follows:
\begin{align}\label{Proof}
    \slashed{D}^2\psi&=\left(\gamma^\mu \gamma^\nu\mathcal{D}_\mu\mathcal{D}_\nu\right)\psi\nonumber\\
    &=\left(\tfrac{1}{2}\{\gamma^\mu,\gamma^\nu\}\mathcal{D}_\mu \mathcal{D}_\nu+\tfrac{1}{2}[\gamma^\mu,\gamma^\nu]\mathcal{D}_\mu\mathcal{D}_\nu\right)\psi\nonumber\\
    &=\left(g^{\mu\nu}\mathcal{D}_\mu \mathcal{D}_\nu+\tfrac{1}{2}[\gamma^\mu,\gamma^\nu]\mathcal{D}_\mu\mathcal{D}_\nu\right)\psi\nonumber\\
    &=\left(g^{\mu\nu}\mathcal{D}_\mu \mathcal{D}_\nu+\tfrac{1}{4}[\gamma^\mu,\gamma^\nu][\mathcal{D}_\mu,\mathcal{D}_\nu]\right)\psi\nonumber\\
    &=\left(g^{\mu\nu}\mathcal{D}_\mu \mathcal{D}_\nu-\tfrac{1}{4}[\gamma^\mu,\gamma^\nu]R_{\mu\nu}^{\,\,\,\,ab}\Sigma_{ab}\right)\psi\nonumber\\
    &=\left(g^{\mu\nu}\mathcal{D}_\mu \mathcal{D}_\nu-2R_{\mu\nu\rho\sigma}\Sigma^{\mu\nu}\Sigma^{\rho\sigma}\right)\psi\nonumber\\
    &=\left(g^{\mu\nu}\mathcal{D}_\mu \mathcal{D}_\nu+\tfrac{1}{4}R\right)\psi.
\end{align}
In the fifth line, we used the following identity:
\begin{equation}\label{OperatorCommutation}
    [\mathcal{D}_\mu,\mathcal{D}_\nu]\psi=-R_{\mu\nu}^{\,\,\,\,ab}\,\Sigma_{ab}\,\psi,
\end{equation}
which\footnote{Note that commutation relation (\ref{OperatorCommutation}), and its generalization to the case where the Maxwell field is present, was first derived by Fock in Ref.\,\cite{Fock1929}.}, itself, follows from the following expression of the Riemann tensor in terms of the spin connection:  
\begin{equation}
    R_{\mu\nu}^{\,\,\,\,ab}=\partial_\nu\omega_{\mu}^{\,ab}-\partial_\mu\omega_{\nu}^{\,ab}+\omega_{\nu}^{\,ac}\omega_{\mu c}^{\,\,\,b}-\omega_{\mu}^{\,ac}\omega_{\nu c}^{\,\,\,b}.
\end{equation}
In the last line of Eq.\,(\ref{Proof}), we used the cyclic property of the Riemann tensor $R_{\mu\nu\rho\sigma}+R_{\mu\rho\sigma\nu}+R_{\mu\sigma\nu\rho}=0$, the anticommutation relations $\{\gamma_\mu,\gamma_\nu\}=2g_{\mu\nu}$ of the gamma matrices, as well as the definition of the Ricci scalar $R=g^{\mu\nu}R_{\mu\nu}=g^{\mu\nu}R^\rho_{\,\,\,\mu\rho\nu}$.
Thanks to Eq.\,(\ref{Proof}), one easily derives the Schr\"odinger-Dirac equation by following the usual steps from the literature by applying from the left the operator $i\gamma^\mu D_\mu+m$ to the Dirac equation $(i\gamma^\mu D_\mu-m)\,\psi=0$:
\begin{align}\label{ProofSchroDirac}
    (i\gamma^\mu D_\mu +m)(i\gamma^\nu D_\nu-m)\,\psi&=0\nonumber\\
    \Longrightarrow\;\left(\slashed{D}^2\psi+m^2\right)\psi&=0\nonumber\\
    \Longrightarrow\;\left(g^{\mu\nu}\mathcal{D}_\mu \mathcal{D}_\nu+\tfrac{1}{4}R+m^2\right)\psi&=0.
\end{align}
It is worth noting here that nowhere along these steps the Dirac equation itself has been used. All one has to assume is that the second-order differential equation in the very first line holds; the Schr\"odinger-Dirac equation in the last line follows then automatically. In fact, if one were to {\it assume} that the Dirac equation holds, then one only needs to apply to the Dirac equation from the left the operator $\gamma^\mu D_\mu$:
\begin{align}\label{SchroDiracWithDirac}
    \gamma^\mu D_\mu(i\gamma^\nu D_\nu-m)\,\psi&=0\nonumber\\
    \Longrightarrow\;\left(g^{\mu\nu}\mathcal{D}_\mu \mathcal{D}_\nu+\tfrac{1}{4}R+im\gamma^\mu D_\mu\right)\psi&=0.
\end{align}
The second line in these steps is indeed a novel second-order differential equation that reduces to the Schr\"odinger-Dirac equation only when the Dirac equation holds. 

\section{Conformal invariance of the curved-spacetime Dirac equation}\label{sec:AppConfDirac}
In this appendix, we first gather the various formulas required in Sec.\,\ref{sec:ConfSchroDirac}, and then we give the detailed derivation of the conformal invariance of the curved-spacetime Dirac  equation.

Under the Weyl transformation of spacetime, $\tilde{g}_{\mu\nu}=\Omega^2g_{\mu\nu}$, the vierbeins $e^a_\mu$, their inverses $e^\mu_a$, the Dirac field $\psi$ and the mass $m$ of the latter are all affected as follows:
\begin{equation}\label{Conformalvierbein}
    e^a_\mu=\Omega^{-1}\tilde{e}^a_\mu,\quad e^\mu_a=\Omega\,\tilde{e}^\mu_a,\quad \psi=\Omega^s\tilde{\psi},\quad m=\Omega\,\tilde{m}.
\end{equation}
Note that we chose here to assign the conventional conformal factor $\Omega^s$ for the spinor field $\psi$ in order to show that the Dirac equation is conformally invariant only for the specific weight $s=\frac{3}{2}$. A different conformal factor for the spinor field is introduced and analyzed in Sec.\,\ref{sec:ConfSchroDirac}.

As a consequence of the transformation of the vierbeins, the Christoffel symbols $\Gamma_{\mu\nu}^\lambda$, the spin connection $\omega_{\mu}^{\,\,ab}$, the spin-covariant derivative $D_{\mu}$ and the Ricci scalar $R$ are all affected as follows:
\begin{align}\label{ConformalConnections}
    \Gamma_{\mu\nu}^\lambda&=\tilde{\Gamma}_{\mu\nu}^\lambda-\Omega^{-1}\left(\tilde{\delta}^\lambda_\mu\Omega_{,\nu}+\tilde{\delta}^\lambda_\nu\Omega_{,\mu}-\tilde{g}_{\mu\nu}\tilde{g}^{\lambda\rho}\Omega_{,\rho}\right),\nonumber\\
    \omega_\mu^{\,\,ab}&=\tilde{\omega}_\mu^{\,\,ab}+\Omega^{-1}\Omega_{,\nu}\left(\tilde{e}^{\nu a}\tilde{e}^b_{\mu}-\tilde{e}^a_\mu\tilde{e}^{\nu b}\right),\nonumber\\
    D_\mu&=\tilde{D}_\mu+2\Omega^{-1}\Omega_{,\nu}\tilde{\Sigma}^\nu_{\,\,\mu},\nonumber\\
    R&=\Omega^2\tilde{R}-6\,\Omega\,\tilde{\Box}\Omega+12\tilde{g}^{\mu\nu}\Omega_{,\mu}\Omega_{,\nu}.
\end{align}
For convenience, we have denoted the partial derivatives $\partial_\mu\Omega$ by $\Omega_{,\mu}$ and we introduced the d'Alembertian in the conformal frame, $\tilde{\Box}=\tilde{\nabla}_\mu\tilde{\nabla}^\mu$. Using these formulas, and starting from the curved-spacetime Dirac equation written in the original frame, we prove the conformal invariance of the equation as follows:
\begin{align}\label{ProofConformalDirac}
(i\gamma^\mu D_\mu-m)\,\psi&=0\nonumber\\
\Rightarrow(i\Omega\tilde{\gamma}^\mu\tilde{D}_\mu+2i\tilde{\gamma}^\mu\Omega_{,\nu}\tilde{\Sigma}^\nu_{\,\,\mu}-\Omega \,\tilde{m})\Omega^s\tilde{\psi}&=0\nonumber\\
\Rightarrow\Omega^{s+1}\left(i\tilde{\gamma}^\mu \tilde{D}_\mu-\tilde{m}+i\left[s-\tfrac{3}{2}\right]\frac{\Omega_{,\mu}}{\Omega}\tilde{\gamma}^\mu\right)\tilde{\psi}&=0.
\end{align}
To go from the third to the fourth step, we applied the partial derivative on the factor $\Omega^s$, and then we used the fact that $2\tilde{\gamma}^\mu\tilde{\Sigma}_{\nu\mu}=-\frac{3}{2}\tilde{\gamma}_\nu$ which is, itself, derived from the usual identities satisfied by the Dirac matrices in curved spacetime; namely $\tilde{\gamma}^\mu\tilde{\gamma}^\nu\tilde{\gamma}_\mu=-2\tilde{\gamma}^\nu$ and $\tilde{\gamma}^\mu\tilde{\gamma}_\mu=4$. The last equation shows that only for $s=\frac{3}{2}$ does the equation become conformally invariant. For a derivation of the conformal invariance of the Dirac equation based on the Dirac Lagrangian in curved spacetime see Refs.\,\cite{QFT1,Conformal}, and for a derivation based on the massless Dirac equation see Ref.\,\cite{QFT2}.

\section{Conformal transformation of the Schr\"odinger-Dirac equation}\label{sec:AppConfSchrodirac}
Using the various conformal transformations as given in Eqs.\,(\ref{Conformalvierbein}) and (\ref{ConformalConnections}) of the previous appendix, and starting from the Schr\"odinger-Dirac equation (\ref{SchroDirac}) written in the original frame, the final form for the equation in the conformal frame is found as follows:
\begin{align}\label{ConfSchroDiracTransformation}
    &\left(g^{\mu\nu}\mathcal{D}_\mu \mathcal{D}_\nu+m^2+\tfrac{1}{4}R\right)\psi=0\nonumber\\
&\Rightarrow\;\left(g^{\mu\nu}D_\mu D_\nu-g^{\mu\nu}\Gamma_{\mu\nu}^\lambda D_\lambda+m^2+\tfrac{1}{4}R\right)\psi=0\nonumber\\
&\Rightarrow\;\Bigg(\Omega^2\tilde{g}^{\mu\nu}\left[\tilde{D}_\mu+2\Omega^{-1}\Omega_{,\rho}\tilde{\Sigma}^\rho_{\,\,\mu}\right] \left[\tilde{D}_\nu+2\Omega^{-1}\Omega_{,\rho}\tilde{\Sigma}^\rho_{\,\,\nu}\right]\nonumber\\
&\qquad-\left[\Omega^2\tilde{g}^{\mu\nu}\tilde{\Gamma}_{\mu\nu}^\lambda+2\Omega\tilde{g}^{\lambda\mu}\Omega_{,\mu}\right]\left[\tilde{D}_\lambda+2\Omega^{-1}\Omega_{,\rho}\tilde{\Sigma}^\rho_{\,\,\lambda}\right]\nonumber\\
&\qquad+\Omega^2\tilde{m}^2+\tfrac{1}{4}\Omega^2\tilde{R}-\tfrac{3}{2}\,\Omega\tilde{\Box}\Omega+3\tilde{g}^{\mu\nu}\Omega_{,\mu}\Omega_{,\nu}\Bigg)\,\Omega^{\frac{3}{2}}\tilde{\psi}=0\nonumber\\
&\Rightarrow\;\Omega^{\frac{7}{2}}\left[\left(\tilde{g}^{\mu\nu}\tilde{\mathcal{D}}_\mu\tilde{\mathcal{D}}_\nu+\tilde{m}^2+\tfrac{1}{4}\tilde{R}\right)\tilde{\psi}+\frac{\Omega_{,\mu}}{\Omega}\left(4\tilde{\Sigma}^{\mu\nu}\!+\tilde{g}^{\mu\nu}\right)\tilde{D}_{\nu}\tilde{\psi}\right]\!=\!0\nonumber\\
&\Rightarrow\;\Omega^{\frac{7}{2}}\left[\left(\tilde{g}^{\mu\nu}\tilde{\mathcal{D}}_\mu\tilde{\mathcal{D}}_\nu+\tilde{m}^2+\tfrac{1}{4}\tilde{R}\right)\tilde{\psi}+\frac{\Omega_{,\mu}}{\Omega}\tilde{\gamma}^\mu\tilde{\gamma}^\nu\tilde{D}_\nu\tilde{\psi}\right]=0\nonumber\\
&\Rightarrow\left(\tilde{g}^{\mu\nu}\tilde{\mathcal{D}}_\mu\tilde{\mathcal{D}}_\nu+\tilde{m}^2+\tfrac{1}{4}\tilde{R}\right)\tilde{\psi}=-\frac{\Omega_{,\mu}}{\Omega}\tilde{\gamma}^\mu\tilde{\gamma}^\nu\tilde{D}_\nu\tilde{\psi}.
\end{align}
In the first step, we expressed the total covariant derivative $\mathcal{D}_\mu$ in terms of $D_\mu$ and the Christoffel symbols. In the second step, we used the identities in Eq.\,(\ref{ConformalConnections}). In the third step, we used the usual identities satisfied by the Dirac gamma matrices in curved spacetime: $\tilde{\gamma}^\mu\tilde{\gamma}_\mu=4$,  $\tilde{\gamma}^\mu\tilde{\gamma}^\nu\tilde{\gamma}_\mu=-2\tilde{\gamma}^\nu$ and $\tilde{\gamma}^\mu\tilde{\gamma}^\nu\tilde{\gamma}^\rho\tilde{\gamma}_\mu=4\tilde{g}^{\nu\rho}$, as well as the anti-symmetry of the spin tensor: $\tilde{\Sigma}^{\mu\nu}=-\tilde{\Sigma}^{\nu\mu}$ and the fact that the latter is covariantly constant: $\tilde{\mathcal{D}}_\nu\tilde{\Sigma}^{\mu\nu}=0$. In the fourth step, we used the identity $4\tilde{\Sigma}^{\mu\nu}=\tilde{\gamma}^{\mu}\tilde{\gamma}^{\nu}-\tilde{g}^{\mu\nu}$.
\section{Conformal transformation of Lichnerowicz's identity}\label{sec:AppConfLichSchro}
Using the various conformal transformations as given in Eqs.\,(\ref{Conformalvierbein}) and (\ref{ConformalConnections}) of the previous appendix, and assuming identity (\ref{Theorem}) holds in the original spin manifold, we derive here the conformal version of the latter identity in the new spacetime.

Let us start by conformally transforming the left-hand side of identity (\ref{Theorem}) as follows:
\begin{align}\label{ConfTheorem1}
    \int{\rm d}^4x\sqrt{g}\,\psi^\dagger\slashed{D}^2\psi&=\int{\rm d}^4x\,\Omega^{-\frac{3}{2}}\sqrt{\tilde{g}}\,\tilde{\psi}^\dagger\tilde{\gamma}^\mu
    \left(\tilde{D}_\mu+2\frac{\Omega_{,\rho}}{\Omega}\tilde{\Sigma}^\rho_{\,\,\mu}\right)\Omega\tilde{\gamma}^\nu\left(\tilde{D}_\nu+2\frac{\Omega_{,\lambda}}{\Omega}\tilde{\Sigma}^\lambda_{\,\,\nu}\right)\!\Omega^{\frac{3}{2}}\tilde{\psi}\nonumber\\
    &=\int{\rm d}^4x\sqrt{\tilde{g}}\,\Omega\,\tilde{\psi}^\dagger\tilde{\slashed{D}}^2\tilde{\psi}+\int{\rm d}^4x\sqrt{\tilde{g}}\,\Omega_{,\mu}\,\tilde{\psi}^\dagger\tilde{\gamma}^\mu\tilde{\slashed{D}}\tilde{\psi}.
\end{align}
In the second step, we expanded the parentheses and used again the fact that $\tilde{\gamma}^\mu\tilde{\Sigma}^\nu_{\,\,\mu}=-\tfrac{3}{4}\tilde{\gamma}^\nu$, which makes the majority of the terms cancel each other. 
On the other hand, using the result (\ref{ConfSchroDiracTransformation}), we find that the right-hand side of the first line of identity (\ref{Theorem}) transforms as follows:
\begin{align}\label{ConfTheorem2}
    &-\int{\rm d}^4x\sqrt{g}\,\psi^\dagger\mathcal{D}^\mu \mathcal{D}_\mu\psi+\tfrac{1}{4}\int{\rm d}^4x\sqrt{g}\,\psi^\dagger R\,\psi\nonumber\\
    &=\!\int\!{\rm d}^4x\!\sqrt{\tilde{g}}\,\Omega\,\tilde{\psi}^\dagger\!\left(-\tilde{g}^{\mu\nu}\tilde{\mathcal{D}}_\mu\tilde{\mathcal{D}}_\nu+\tfrac{1}{4}\tilde{R}\right)\tilde{\psi}+\!\!\int\!{\rm d}^4x\!\sqrt{\tilde{g}}\,\Omega_{,\mu}\,\tilde{\psi}^\dagger\tilde{\gamma}^\mu\tilde{\slashed{D}}\tilde{\psi}\nonumber\\
    &=\!\int{\rm d}^4x\sqrt{\tilde{g}}\,\Omega\,|\tilde{D}\,\tilde{\psi}|^2+\tfrac{1}{4}\!\int{\rm d}^4x\sqrt{\tilde{g}}\,\Omega\,\tilde{R}\,|\tilde{\psi}|^2+\!\int\!{\rm d}^4x\sqrt{\tilde{g}}\,\Omega_{,\mu}\tilde{\psi}^\dagger\,\tilde{g}^{\mu\nu}\tilde{D}_\nu\tilde{\psi}+\!\int\!{\rm d}^4x\sqrt{\tilde{g}}\,\Omega_{,\mu}\,\tilde{\psi}^\dagger\tilde{\gamma}^\mu\tilde{\slashed{D}}\tilde{\psi}.
\end{align}
In the second step, we integrated by parts. Equating the right-hand sides of the results (\ref{ConfTheorem1}) and (\ref{ConfTheorem2}), and then using the identity $4\tilde{\Sigma}^{\mu\nu}=\tilde{\gamma}^{\mu}\tilde{\gamma}^{\nu}-\tilde{g}^{\mu\nu}$, we arrive at the following conformal transformation of identity (\ref{Theorem}):
\begin{align}
\!\int{\rm d}^4x\!\sqrt{\tilde{g}}\,\Omega\,\tilde{\psi}^\dagger\tilde{\slashed{D}}^2\tilde{\psi}&=\!\int\!{\rm d}^4x\sqrt{\tilde{g}}\,\Omega\,|\tilde{D}\,\tilde{\psi}|^2+\tfrac{1}{4}\!\int\!{\rm d}^4x\!\sqrt{\tilde{g}}\,\Omega\,\tilde{R}\,|\tilde{\psi}|^2+\int{\rm d}^4x\sqrt{\tilde{g}}\,\Omega^{,\mu}\tilde{\psi}^\dagger\tilde{D}_\mu\tilde{\psi}.
\end{align}


\end{document}